\documentclass[lettersize,journal]{IEEEtran}

\ifCLASSINFOpdf
\else
\fi
\ifCLASSOPTIONcompsoc
  \usepackage[nocompress]{cite}
\else
  \usepackage{cite}
\fi

\usepackage{xurl}

\usepackage{graphics}
\usepackage{amsmath}
\usepackage{listings}
\usepackage{graphicx}

\usepackage{support-caption}
\usepackage{subcaption}
\usepackage{lipsum}

\usepackage[table,xcdraw]{xcolor}
\usepackage{tikz}
\usepackage{tikz-dependency}
\newcommand*\circled[1]{\tikz[baseline=(char.base)]{        
 \node[shape=circle,fill,inner sep=0.5pt] (char) {\textcolor{white}{#1}};}}
\usepackage{comment}
\usepackage{amssymb}% http://ctan.org/pkg/amssymb
\usepackage{pifont}% http://ctan.org/pkg/pifont

\begin{document}
%\title{Root Cause Analysis for 5G Network Failures}

%\title{Simba: A Graph Neural Network and Transformer-based Anomaly Detector and Root Cause Analyzer in 5G
\title{Root Cause Analysis of Anomalies in 5G RAN Using Graph Neural Network and Transformer
\thanks{$^\circ$ These student authors contributed equally as first authors.}}

\author{
 \IEEEauthorblockN{Antor Hasan$^\star$$^\circ$, Conrado Boeira$^\star$$^\circ$, Khaleda Papry$^\star$$^\circ$, Yue Ju$^\diamond$, Zhongwen Zhu$^\diamond$, Israat Haque$^\star$}
 
\IEEEauthorblockA{$^\diamond$Ericsson GAIA, Montreal, Canada}
\IEEEauthorblockA{$^\star$Department of Computer Science, Dalhousie University, Canada}
}

\maketitle


\begin{abstract}

The emergence of 5G technology marks a significant milestone in developing telecommunication networks, enabling exciting new applications such as augmented reality and self-driving vehicles. However, these improvements bring an increased management complexity and a special concern in dealing with failures, as the applications 5G intends to support heavily rely on high network performance and low latency. Thus, automatic self-healing solutions have become effective in dealing with this requirement, allowing a learning-based system to automatically detect anomalies and perform Root Cause Analysis (RCA). However, there are inherent challenges to the implementation of such intelligent systems. First, there is a lack of suitable data for anomaly detection and RCA, as labelled data for failure scenarios is uncommon. Secondly, current intelligent solutions are tailored to LTE networks and do not fully capture the spatio-temporal characteristics present in the data. Considering this, we utilize a calibrated simulator, Simu5G, and generate open-source data for normal and failure scenarios. Using this data, we propose \textit{Simba}, a state-of-the-art approach for anomaly detection and root cause analysis in 5G Radio Access Networks (RANs). We leverage Graph Neural Networks to capture spatial relationships while a Transformer model is used to learn the temporal dependencies of the data. We implement a prototype of \textit{Simba} and evaluate it over multiple failures. The outcomes are compared against existing solutions to confirm the superiority of \textit{Simba}. 

\end{abstract}

% \fi
\IEEEpeerreviewmaketitle

%\vspace{-5mm}
\section{Introduction}
\label{sec:introduction}

The advent of 5G stands out as a pivotal development in the contemporary technological landscape, playing an essential role in shaping the future of telecommunications. Leveraging new functionalities such as network slicing and mmWave frequency bands, 5G offers a new era of mobile networks. These advancements enhance the overall performance of networks and constitute the foundation for a wide range of cutting-edge applications like intelligent transportation, smart health, and autonomous vehicles. However, these improvements come at the cost of increased management complexity of the 5G ecosystem. Estimates show the management expenditure of providers is around 10\%-30\% of their revenue \cite{DeGrasse_2017}. Moreover, the above latency-critical applications demand the network to be highly reliable, as any disruption or anomalies will lead to subpar application performance. Self-organization networks (SON) have gained the attention of providers and researchers to automatically detect and prevent failures or anomalies in 5G networks. SON is composed of three components: self-configuration, self-optimization, and self-healing \cite{aliu2012survey}. 

Self-healing can bring the needed reliability to the applications described above. Specifically, it refers to learning-based anomaly detection and root cause analysis (RCA) to realize resilient 5G networks for latency-critical applications \cite{moysen20184g}. Identifying and categorizing failures in 5G networks is challenging as it depends significantly on spatial and temporal characteristics of the network, which dynamically change over time due to user mobility, i.e., adds another layer of complexity to RCA. Specifically, the key performance indicators (KPIs) of communications are highly dynamic and have spatial and temporal characteristics. 
Moreover, failures are not isolated phenomena that affect only the suffering base station, instead its neighboring base stations may suffer from a performance degradation \cite{gomez2015methodology}. For example, a base station suffering from a transmission power reduction will have reduced coverage, forcing neighboring base stations to extend their coverage.

%\textcolor{blue}{Root-Cause Analysis, i.e., detecting and categorizing failures that happen in the network, can be a specially challenging problem in telecommunication networks as it depends greatly on temporal and spatial features of the network. As users are continuously moving and communicating, measurements from base-stations are in constant change, and drops and spikes in performance are not uncommon and do not necessarily point to a failure in the network. This characteristic makes it vital to consider the temporal aspect of it when making predictions. Moreover, failures are not isolated phenomena that affect only the cell that is causing it. Neighboring cells might suffer a degradation in performance as well. For example, a cell suffering from reduction in it's transmission power might see a reduction in it's coverage, forcing neighboring cells to serve users further than they normally would, reducing the signal strength they can provide.}

%\textcolor{red}{state why learning-based anomaly detection and RCA is useful in this complex 5G networks.}
Learning-based anomaly detection and RCA is an inevitable choice due to the availability of a large volume of 5G data (e.g., KPIs across network elements) and the complex nature of the 5G architecture and its scale. Also, these automated solutions are beneficial in reducing human errors, Operating Expenses (OPEX), and Capital Expenditures (CAPEX). Providers can efficiently detect and correct anomalies to meet the demand of emerging applications.  However, existing learning-based RCA schemes are mostly for LTE networks \cite{zhang2019self,chen2020active,mfula2017adaptive}, which incorporate KPIs of LTE base stations to detect anomalies. Yen et al. \cite{yen2022graph} proposed a Graph Neural Network (GNN) model for detecting cell outages in 5G radio access networks (RANs). In addition to cell outage, other failures or anomalies happen in 5G RAN, e.g., Excessive Antenna Uptilt and Downtilt, Inter-system Interference, Excessive Power Reduction (EPR), and Too-Late Handover (TLHO) \cite{gomez2015methodology,chen2020active}. 
Out of these anomalies, we choose the most common failures in 5G \cite{tarrias2023failure}, namely EPR and Inter-System Interference. 

In designing learning-based RCA, we identified two critical gaps that existing works failed to fill. First, existing solutions failed to provide detailed description of their used data and anomalies, which limit their contribution. In particular, as it is difficult to reproduce failed events, which is smaller in percentage compared to normal behavior, we must use either real failed data or realistically generated synthetic anomalies. Existing work failed to incorporate either scheme. Second, the state-of-the-art solution in 5G do not incorporate the spatio-temporal correlations among failure events while developing learning-based models. Specifically, Yen et al.\cite{yen2022graph} developed a Graph Neural Network model to capture spatial context of KPIs in detecting cell outages without incorporating the temporal context. However, the KPIs have the spatio-temporal characteristics that must be captured in the developed model. For example, MTGNN \cite{wu2020connecting} is a weather forecasting model based on the spatio-temporal weather patterns, where the temporal pattern is captured using 1D convolution networks. Thus, we first modified MTGNN in predicting anomalies and performing RCA. MTGNN architecture does not incorporate the state-of-the-art time series analyzer Transformer \cite{vaswani2017attention} despite a better performance than using only GNNs.

We present, \textit{Simba}, an anomaly detection and RCA scheme that combines  GNN and the state-of-the-art time series analyzer Transformer. In particular, the model has three main components: a Graph Learning (GL) module, a Graph Convolution Module (GCN) and a Transformer module. The former learns the graph representation of the base stations, which is then fed to the GCN module to learn their spatial embedding. On the other hand, the Transformer module learns the temporal embedding of the base stations. The GCN module then learns the combined spatial-temporal embedding for the anomaly and RCA prediction. Specifically, a Feed Forward Network (FNN) generates the probability distribution of the different anomalies as an output vector.

In addition to identifying the anomalies, Simba also locates the base stations that suffer from such disputes. We generate anomalies and regular operational data in a realistic 5G simulator, Simu5G \cite{nardini2020simu5g}, calibrated following 3GPP specifications in collaboration with our industrial partner, Ericsson \cite{boeira2024calibrated,2024calibrating}. The data generation procedure, with adequate descriptions, is presented in our GitRepo \cite{repo}. Then, we perform an extensive evaluation of the proposed scheme to show its effectiveness. Unlike existing works, we consider multiple simultaneous anomalies and effectively detect them using the proposed model. The results demonstrate that \textit{Simba} outperforms the state-of-the-art solutions and show substantial improvement in accurately identifying, classifying, and locating failures happening in a network. The main contributions of this work are as follows:

\begin{itemize}
    \item This work proposes a novel architecture of a time-series Transformer with GNN spatial embedding for detecting and locating failures in 5G RANs, achieving high performance (around 0.8 F1 score for multiple failures).
    \item To the best of our knowledge, this work is the first to propose an RCA model targeting 5G networks that can detect and locate simultaneous failures in 5G networks.
    \item We thoroughly evaluate the performance of existing models and compare them to the proposed one in a couple of scenarios.  %with single and multiple failures.
    \item We share the developed model and the generated data over a Git Repository \cite{repo} for reproducibility and extension. 
\end{itemize}

\section{Background and Motivation}
\label{background}

This section first presents the necessary background on the chosen 5G faults. Then, we introduce the main building blocks of the proposed model, i.e., GNN, GCN, and Transformer, with the rationale of choosing them.

\textbf{Failures in 5G deployments.}
There are multiple potential failures in Radio Access Networks, ranging from holes in the coverage offered by a provider to improper positioning of antennas \cite{gomez2015methodology,zhang2019self,chen2020active}. In this work, we focus on two of the most common faults as pointed by our industry partner: \textit{excessive power reduction} and \textit{interference}

\textit{Excessive power reduction}: occurs when the output power of a cell diminishes to a point where it is not able to meet the necessary level of expected performance. For example, urban cells operating with 10MHz of bandwidth are expected to have a transmission power of 41 to 45 dBm \cite{ituimt,chen2020active}, a significant drop from this value can lead to narrow coverage and poor signal strength for the cell. This problem usually arise from sub-optimal parameter configurations or issues in the wiring system \cite{gomez2015methodology}.

\textit{Inter-cell Interference}: stands out as one of the main issues causing disruption in mobile networks. This phenomenon appears when users from neighboring cells operate on the same frequency, which leads to mixing between the intended receiving and interfering signals, resulting in reception distortion \cite{siddiqui2021interference}. This failure can occur when users from other cells utilize the same frequency or these neighboring base stations are configured incorrectly  
% \textit{Inter-Cell Interference}: is one significant cause of issues in telecommunication networks, and it happens when users of neighbor cells use the same frequency. Thus, the intended receiving signals are mixed with the interfering signals and distorted the reception \cite{siddiqui2021interference}. Such interference can occur if users from the neighboring cells use the same frequency or those cells are incorrectly configured. 
%\subsection{AI/ML Approaches }

\begin{figure}[h] 
    \centering
    \includegraphics[width=.48\textwidth]{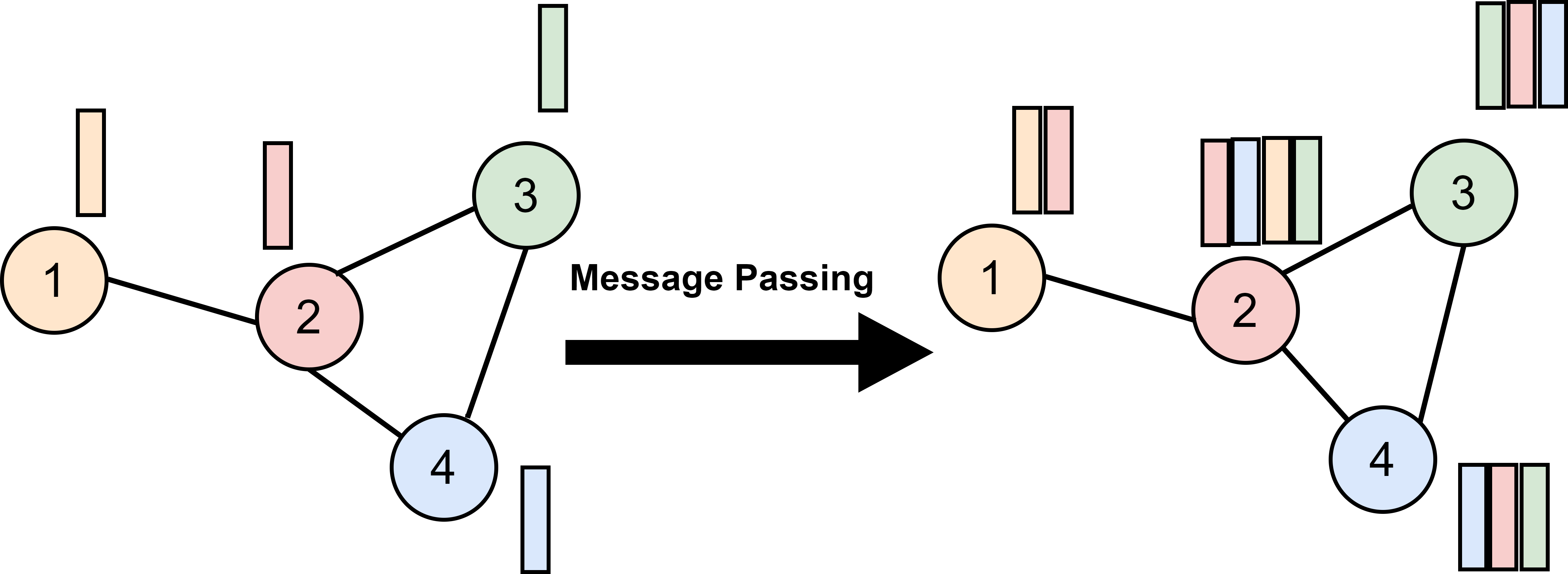}
    \caption{Message passing of a GNN.}
    \label{fig:gnn}
\end{figure}

\textbf{Graph neural networks.} 
A graph neural network (GNN) is a class of artificial neural networks for processing data that can be represented as graphs, e.g., social networks, knowledge graphs, recommender systems, and bioinformatics. Usually, nodes from a graph use message passing to construct their edges to represent their adjacency relationship. Specifically, each node creates a feature vector and sends it to its neighbors. Thus, a node receives one message per adjacent node. The message passing (Fig.~\ref{fig:gnn}) continues in a recursive or convolutional constructive manner until no further message passing happens and all nodes have their updated message from neighbors. Each node updates its representation as node features by aggregating the messages received from the neighbours and its own features. The GNN model then maps the graph of node features to a multidimensional Euclidean space for each node \cite{scarselli2008graph}. Generally, multiple non-linear transformations or local/global pooling layers are applied to the multidimensional vectors to provide a fixed-size representation of the entire graph. The mapped representation is used for downstream classification or regression (e.g., graph-based, node-based, or edge-based) problem \cite{wu2020comprehensive}. In the proposed problem, we can construct a graph using BSs, mobile users, and their communications, which has a spatial context that dynamically changes over time. Thus, it is viable to use GNN-based learning to capture anomalies. 

%\textcolor{red}{ GNNs are specially used for analyzing graph structured data. A graph exists in non-euclidean space which makes it harder to interpret the data using conventional machine learning approaches. Moreover, graphs are dynamic and do not have a fixed form with possibly much higher dimensions which makes it difficult to analyze data Therefore, it is convenient to use GNN model for non-euclidean, high dimensional, and dynamic data. }
%\textcolor{red}{why part is missing- why we need GNN in our work?}
%Graph-based data structures have drawbacks, and data scientists must understand them before developing graph-based solutions.

%\begin{comment}
%\end{comment}

%\textcolor{red}{Thus, a node receives one message per adjacent node. The message passing continues in a recursive or convolutional constructive manner until \textcolor{red}{a stable fixed point is reached. -- what is the stable point? all nodes have their updated feature vectors and no more changes happen? can you be a bit specific?} The GNN model then maps the graph of node features to a multidimensional Euclidean space for each node \cite{scarselli2008graph}.  These multidimensinal Euclidean spaces can be used for downstream classification or regression. Specifically, Generally, multiple non-linear transformation or local/global pooling layers are applied on the representation to provide a fixed-size representation of the whole graph which are then can be used for different classification (e.g node-based, graph-based or edge-based) problem \cite{wu2020comprehensive}.} 

\begin{figure} [!ht]
    \centering
    \includegraphics[width=.5\textwidth]{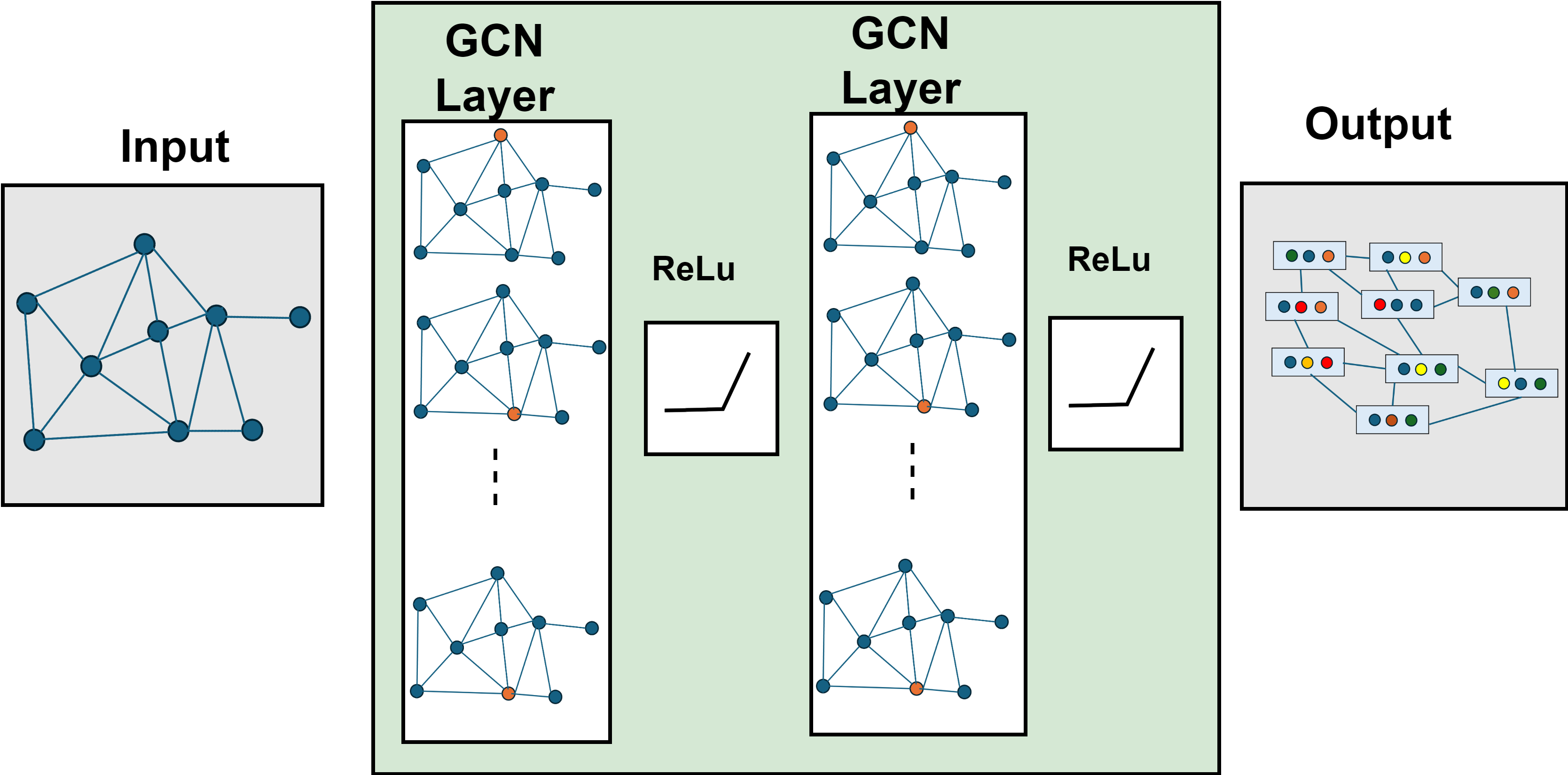}
    \caption{A Graph Convolutional Neural network.}
    \label{fig:gcn}
\end{figure}

\textbf{Graph Convolutional Neural Networks.} GCNs \cite{kipf2016semi} are a type of GNN designed to process graph-structured data with convolutional layers. This is a generalization of convolutional neural network (CNN) for graph structured data. In the first layer of a GCN, a linear transformation is used to transform the input features of each node into a low-dimensional representation. Next, a graph convolution operation is performed to update the representation of each node by combining the representations of its neighbors. The aggregation is done using a weighted sum of the neighbor representations, and the weights are neural layer weights learned during training. A non-linear activation function is applied to the output of the convolutional operation to introduce non-linearity. GCNs can have multiple convolutional layers, allowing them to learn increasingly complex representations of the graph structure. Fig.~\ref{fig:gcn} represents a GCN with multiple convolution layers. Given the previous features, $H^{(l)}$, of nodes the GCN layer is defined as follows:

\begin{equation} \label{eq:gcn}
H^{(l+1)} = \sigma(\hat{D}^{(-1/2)}\hat{A}\hat{D}^{(-1/2)}H^{(l)}W^{(l)})
\end{equation}
Here, $W^{(l)}$  is the weight parameters to transform the input features into messages $( H^{(l)}W^{(l)} )$. In the adjacency matrix $A$, an identity matrix is added, $\hat{A}=A+I$, for each node to send a message to itself. $\hat{D}$ is a diagonal matrix, where $D_{ii}$ is the number of neighbors of node $i$.  $\sigma$  represents an arbitrary activation function, e.g., ReLU is the common activation function in GCNs. As GCN has a better efficiency and generalizability over other GNNs, GCN is being adopted in anomaly detection in 5G \cite{yen2022graph}.

%\textcolor{red}{ The popularity of GCNs has been rapidly growing due to its efficiency and generality in capturing spatial features and convenience to composite with other neural networks. Therefore, it is widely been used in different prediction and classification problems for retrieving spatial dependencies among nodes. }
%\textcolor{blue}{ Based on the applications with GCN model like classification, multiple pooling layers and/ or MLP layers are applied after graph convolutional layers to coarsen a graph into subgraphs and find out the  hidden representations of subgraphs.}

\textbf{Transformer.} It has been widely used in sequence modeling due to the capability of the multihead attention mechanism in long-distance dependencies capturing \cite{vaswani2017attention}. Transformer has an encoder-decoder structure where the encoder maps an sequence of input symbols (input features) and the decoder then generates an output sequence symbols one element at a time. At each step, the model consumes the previously generated symbols as additional input when generating the next step symbols. The model relies entirely on its self-attention mechanism to compute representations of its input and output without using sequence aligned RNNs or convolution. It uses multi-head attention that allows the model to jointly attend to information from different representation sub-spaces at different positions \cite{vaswani2017attention}. 

\begin{figure} [!ht]
    \centering
    \includegraphics[width=.45\textwidth]{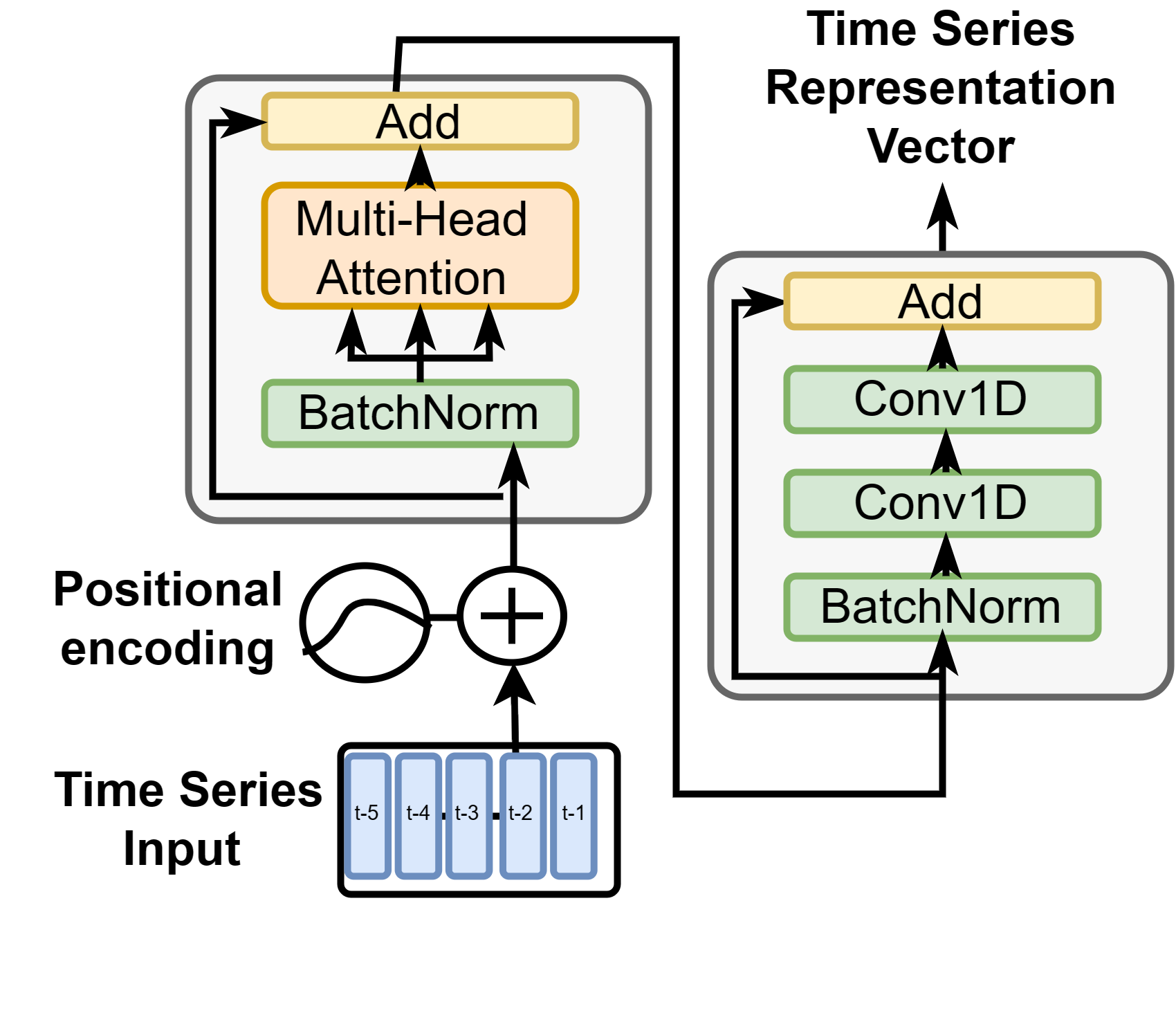}
    %\includesvg[width=0.4\textwidth]{pictures/automation-workflow.drawio.svg}
    \vspace{-.5cm}
    \caption{Transformer module for time-series learning.}
    \label{fig:trans}
\end{figure}

Fig.~\ref{fig:trans} presents the architecture of a time-series transformer with two sub-modules. The first operates on the time series input sequence and performs batch normalization across the feature dimension. A positional encoding is applied before batch normalization to get the relative positions of the input sequence. Then, a multi-head attention mechanism jointly attends to information from different representation sub-spaces at different positions \cite{vaswani2017attention}. The second sub-module has one batch normalization layer along with two 1D convolution layers with Relu activation in between that extracts local patterns and features of the sequence. Residual connections around each of the two sub-modules prevent the vanishing gradient issues. The output time series representation vectors are the sequence that can be fed to downstream tasks like classification and regression.
Transformers can better capture long-term dependencies in time series data compared to CNN and LSTM-based models \cite{zhang2019self, he2020fault}. In addition to the spatial properties, our graph structured data has temporal context, which can be efficiently captured using transformers. 
%\textcolor{red}{Transformers has capability of capturing long distance dependencies among time series data. It can effectively capture temporal dependencies which made it attractive for time series modeling in various time series applications like event predictions and forecasting. As in our model we are using time-series KPI data, transformer is a good choice for handling temporal features of the data. }

\section{Related Work}
\label{sec:related-work}

This section presents the related work of Simba that we group into the following categories. 

\textbf{AI/ML-based RCA.} 
A set of learning-based approaches has been proposed for RCA automation in cellular networks. Chen \textit{et al.} proposed a pool-based active learning and XGBOOST model for fault diagnosis in LTE systems without considering any spatio-temporal data relationship \cite{chen2020root}. Zhang \textit{et al.} filled that gap and combined Convolutional Neural Network (CNN) and Long Short-Term Memory (LSTM) to handle the spatio-temporal dependencies of fault prediction problems in LTE RAN networks \cite{zhang2019self}. In 5G RAN, an ensemble-based LSTM model for multiple faults is proposed in \cite{sundqvist2020boosted}. A decision tree-based failure recovery model for 5G networks is proposed in \cite{omar2021novel}. Multilayer Perceptron, Decision Tree, and Support Vector Machine are used to detect anomaly in 5G Open RAN (O-RAN) in \cite{alves2023machine}. A procedural-based anomaly detection model in 5G RAN is proposed in \cite{sundqvist2023robust}. 
%\textcolor{blue}{ However, all the above mentioned methods needs expert involvement and knowledge to correctly identify root causes. Additionally, the traditional AL/ML methods are not good for handling non-Euclidean structures or irregular data structures like telecommunication network data where GNN-based solutions can be a promising way.}

\textbf{Graph Neural Network-based RCA.} GNN framework for device fault diagnosis in telecommunication networks is proposed in \cite{he2020fault}. Long Short-Term Memory (LSTM) is used initially to get prior information for constructing the GNN. Then, an embedded representation of the graph network is introduced to aggregate the information for the entire graph. Based on the embedded data, it classifies faulty or working devices of the network. MTGNN \cite{wu2020connecting} is proposed for traffic prediction of multivariate time series data, which is used in detecting anomalies from 5G RANs in \cite{yen2022graph}. However, the solution does not handle temporal dependencies among base stations and can only detect anomalies without the RCA. 

\textbf{Transformer-based RCA.} Chen \textit{et al.} proposed a framework for multivariate time-series anomaly detection using transformer \cite{chen2021learning}. They introduce influence propagation convolution to present the anomaly information flow between network nodes. A multi-branch attention mechanism is also applied in the transformer to reduce the model's time complexity. The global-local spatial-temporal transformer network (GLSTTN) \cite{gu2023spatial} performs cellular traffic prediction. The global spatial-temporal module captures global correlations of features using stacked spatial-temporal blocks. In contrast, the local one extracts the local spatial-temporal dependencies hidden in globally encoded features using densely connected CNNs. DetectorNet \cite{li2021detectornet} is a transformer-enhanced GNN for traffic prediction consisting of multi-view temporal attention and a dynamic attention module to capture the long-distance and short-distance temporal correlation and the dynamic spatial correlation of features. Transformer Anomaly \cite{xu2021anomaly} combined transformer and Gaussian prior association to make anomalies more distinguishable. Similarly, TranAD \cite{tuli2022tranad} and BTAD \cite{ma2023btad} are transformer-based anomaly detection for time-series data. Despite the usage of transformer-based models in time-series data analysis, no work leveraged transformers in 5G RCA.

\textbf{Comparison with existing work.} Traditional AI/ML models are not suitable for RCA in 5G networks due to the lack of efficiency in processing unstructured and irregular data. GNN-based models perform better for such data handling but fail to capture temporal context efficiently. Transformers can fill that gap by capturing the temporal context. However, no work combines GNN and Transformers for efficient spatio-temporal correlation capturing while processing unstructured and irregular 5G data for RCA. The proposed neural network model, Simba, takes the benefit of GNN for its ability to capture the spatial context and transformer for its efficient temporal context capturing and develop a novel anomaly detection and RCA model in 5G RAN. 

\section{Data Collection}
\label{data-collection}

This section presents the data that we generated to develop the proposed learning-based model. 

Realistic data is the cornerstone of building any learning-based model like the proposed one. Specifically, Simba needs sufficient data samples (e.g., KPIs across base stations) from normal and anomalies to learn their distribution. Unfortunately, there are no such available datasets as providers do not share such data with ground truth due to privacy and intellectual property concerns. Also, it may not be feasible to label anomalies and log them as they occur unpredictably in a randomly distributed occurrence pattern \cite{chen2020root}. Thus, we chose a simulator with a realistic data generation capability, e.g., Simu5G. Specifically, it is calibrated following 3GPP specifications \cite{ituimt} and can generate all the chosen faults along with normal operational data samples for varying sizes of 5G RANs.  

% \begin{figure} [!ht]
% \centering
% \begin{subfigure}[c]{0.55\columnwidth}
%   \centering
%   \includegraphics[width=0.95\linewidth]{figs/chicago-deployment.drawio.png}
%   \caption{\vspace{-6.25mm}Chicago downtown deployment.}
%   \label{fig:yaml}
% \end{subfigure}%
% \begin{subfigure}[c]{0.45\columnwidth}
%   \centering
%   \includegraphics[width=0.95\linewidth]{figs/hexagonal-deployment.drawio.png}
%   \caption{3GPP recommended deployment.}
%   \label{fig:ini}
% \end{subfigure}
% \caption{Deployment scenarios for data generation.}
% \label{fig:topo}
% \end{figure}

\begin{figure} [!ht]
    \centering
    \includegraphics[width=0.5\columnwidth]{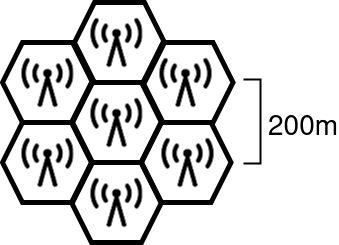}
    \caption{Deployment scenario for data generation}
    \label{fig:topo}
\end{figure}

%We consider two deployment scenarios: AT\&T deployment in downtown Chicago \cite{narayanan2022comparative} and 3GPP recommended deployment \cite{ituimt,zhang2019self}. In the former one, there are 5 base stations deployed on a rectangular area $385 \times 130$ meters. In the 3GPP deployment,  there are a total of 7 cell sites placed in a hexagonal shape, each base station separated by 200 meters from their neighbor in an urban deployment. We present both deployments in Fig.~\ref{fig:topo} along with the complete list of parameters and settings in Table \ref{tab:simu-setup}.  

For generating the data, we utilized the 3GPP recommended eMBB-Urban scenario \cite{ituimt,zhang2019self}. In this deployment, there are a total of 7 cell sites placed in a hexagonal shape, each base station separated by 200 meters from their neighbor in an urban deployment. We present the deployment in Fig.~\ref{fig:topo} along with the complete list of parameters and settings in Table \ref{tab:simu-setup}.

%\textcolor{red}{did we also consider rural deployment? if yes, we msut add that details.} \textcolor{blue}{we didn't consider rural}
%\textcolor{red}{provide some details like total deployment area, number of base stations, and other relevant information etc.}

{\def\arraystretch{1.4}
\begin{table}[!ht]
    \centering
    \begin{tabular}{|p{2.5cm}|p{5cm}|}
         \hline
         \textbf{Parameters} & \textbf{Values}  \\
         \hline
         Topology & 3GPP Standard\\
         \hline
         Carrier Frequency & 4 GHz\\ \hline
         Bandwidth & 10 MHz\\ \hline
         Antenna & Downtilt: 90\textdegree \newline Azimuth beamwdith: 65\textdegree \newline Elevation beamwidth 65\textdegree \\ \hline
         Channel Model & 3GPP RMa TR 38.901 \cite{3gpp.138.901}\\ \hline
         Handover Algorithm & A3 RSRP, Hysteresis = 0\\ \hline
         Tx Power & 41 dB\\ \hline
         Mobility Model & Random Waypoint Model \\ \hline
         Traffic Model & Full Buffer\\ 
         \hline
    \end{tabular}
    \caption{The list of parameters used in the data generation process.}
    \label{tab:simu-setup}
\end{table}}

%\textcolor{red}{review the following para carefully for any missing information.} \textcolor{purple}{Left comments in purple to differentiate. We just need to remove mention to off-peak hours, as we did only normal and peak scenarios - Conrado}\\

In the above deployment scenario, we consider two user densities per base station to represent the regular and peak hours of 5G RAN operations. Specifically, following our partner's guideline, we choose 30 and 90 users per base station to represent normal and peak hours of operation, respectively \cite{ituimt}. Note that all other parameters remain the same except the user density in these operational scenarios. In the chosen deployments, we implement the faults following the 3GPP and partner's guidelines along with the guidelines from existing literature \cite{chen2020root,zhang2019self,gomez2015methodology}. 

In particular, we reduce the transmission power of a base station below the standard to mimic the Excessive Power Reduction (EPR) fault. 
%In the case of too-late handover (TLHO), we substantially increase the value of the Hysteresis of a chosen base station. 
Futhermore, we alter the value of a transmission angle and its angular attenuation to create substantial interference. The detailed settings and parameters of normal and faulty cases are listed in Table \ref{tab:failures}, which we have chosen following the 3GPP standard and partner's guidelines.       

{\def\arraystretch{1.6}
\begin{table}[!ht]
    \centering
    \begin{tabular}{|m{4cm}|m{4cm}|}
         \hline
         \textbf{Faults} & \textbf{Parameters} \\
         \hline
         Excessive Power Reduction & Tx Power: 10 dB \\
         %\hline
         %Too-Late Handover & Hysteresis: 8 \\
         \hline
         Interference & Tx Power: 33 \newline Downtilt: 15\textdegree \newline Azimuth beamwdith: 70\textdegree \newline Elevation beamwidth : 10\textdegree  \\
         \hline
    \end{tabular}
    \caption{Parameter configuration for faulty base stations.}
    \label{tab:failures}
\end{table}}

The above faults can occur at any time during cellular network operation and impact ongoing services. However, the percentage of such faults or anomalies is significantly lower than normal events. As per our partner and existing literature, the percentage of these anomalies occurrence is at most 2\% of the total operation time of base stations \cite{zhang2019self}. Finally, we must collect the necessary KPIs to detect and predict anomalies. We follow the 3GPP standard, existing literature, and partner guidelines and select a list of KPIs that we collect over time. The list includes RSRP, SINR, RSRQ, Throughput, distance between users and BSs, and users and BSs coordinates and IDs. Specifically, RSRP measures the received signal strength, RSRQ measures the signal quality, and SINR is the ratio of the signal strength and the noise plus interference. The throughput at users is the number of downlink received bits per second. 
%\textcolor{blue}{signal measurement KPIs, Throughput, relative distance between users and the base station they are attached to and individual IDs for both users and base stations. } 

%\textcolor{red}{write the data collection procedure as per our discussion.} \\
\textbf{Data collection procedure.} Once the setup is complete, we trigger user and base station operations for a total period of one hour. Specifically, users are uniformly randomly generated for each base station following the chosen operational density. Then, we randomly select an available base station and a failure at a chosen time and change the parameters to create the chosen fault. This failure state remains for 30 to 40 seconds, after which the station is restored to normal behavior. 

%\textcolor{red}{how do we generate traffic at userd and BSs? how do we collect data? at which interval? how do we aggregate? what's the total number of samples: normal and faulty?} 
We utilize constant downlink traffic from the base station to the user and collect KPI data every time a user receives communication signals. We then aggregate this data in intervals of one second. In addition, as these KPIs are collected on the user side, we average out the data of all users connected to a particular base station to generate the mean of each KPI for every base station. Therefore, we have 3600 events per base station, where around 2\% of them, about 72, correspond to failures.

We generate time series KPIs for both normal and anomalous using the calibrated Simu5G. Specifically, we consider excessive power reduction and interference. The percentage of failures or anomalies is kept at most 2\% following our partner's recommendation, which also aligns with other industrial recommendations. We label the failed data while generating the KPIs. Our data set consists of around three million KPI data points, which include UE positions, serving cell distance, reference signal received power (RSRP), reference signal received quality (RSRQ), signal-to-interference-plus-noise ratio (SINR), and throughput.

\section{Simba Model Development}

This section presents the proposed GNN-Transformer-based (Simba) anomaly prediction and RCA model. We present the workflow of our proposed model, data pre-processing steps and the architecture details of the model. In the following, we elaborate each of the components. 

\subsection{Workflow of Simba}

\begin{figure} [ht]
    \centering
    \includegraphics[width=\columnwidth]{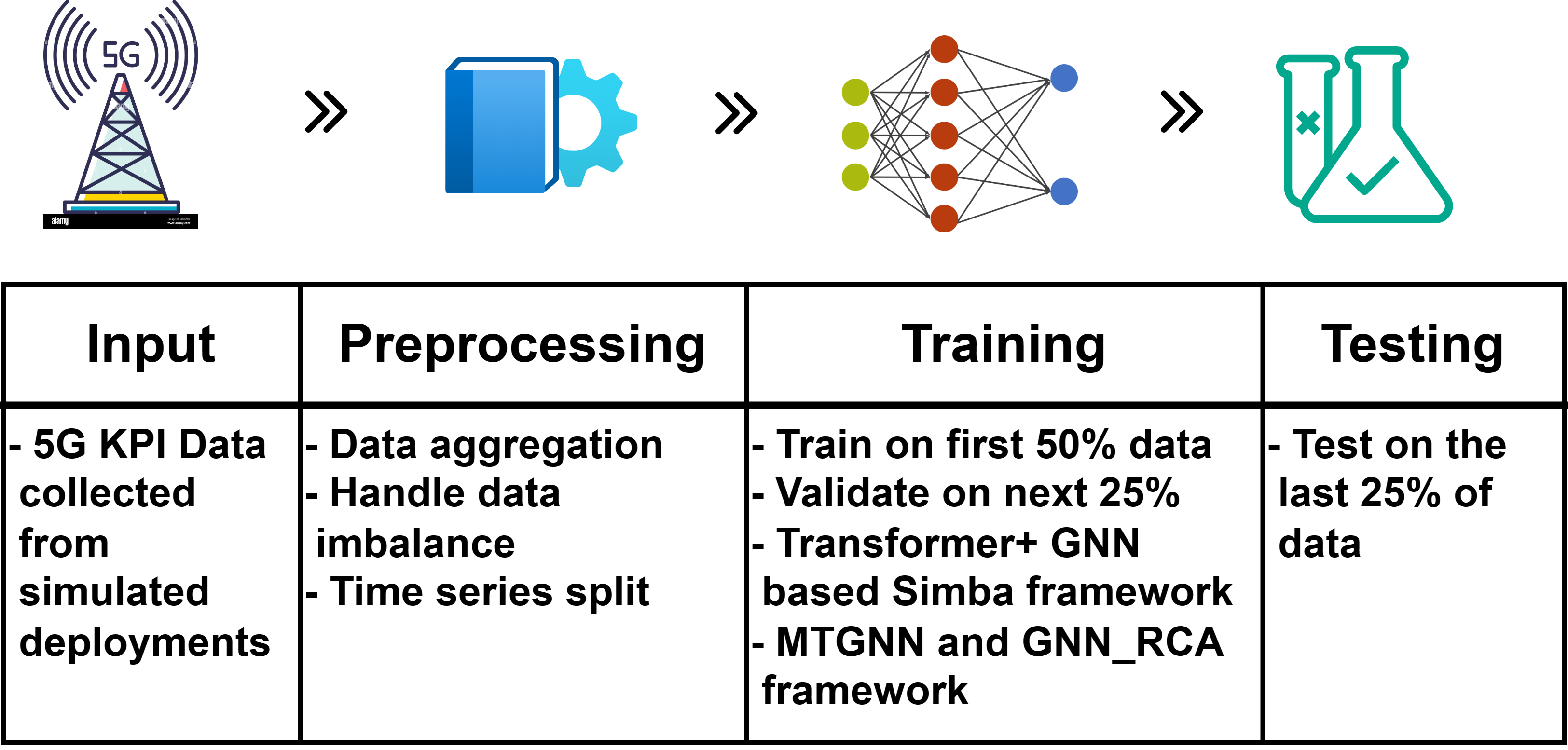}
    \caption{ The workflow of Simba to perform anomaly prediction and RCA in 5G RANs. }
    \label{fig:workflow}
\end{figure}

Fig.~\ref{fig:workflow} presents the overall workflow of our proposed model. It consists of four steps: data collection, data preprocessing, model training and validation, and model testing. Specifically, KPIs are collected from the urban deployment mentioned in Section~\ref{data-collection} with peak and normal operations. Then, these data are pre-processed to make the data suitable for feeding into our model. Data preprocessing consists of data aggregation, imbalance data management, and time series split. Next, we focus on training and validating the proposed GNN-Transformer based model, Sibma. It consists of a GNN module with Graph Structure Learning (GSL) and Graph Convolution (GC) sub modules, a generalized transformer branch, and feed-forward branches. We divide the entire datasets into: the first 50\% for training, the next 25\% for validation, and the last 25\% to test. 
%\textcolor{blue}{We need to change the text in the image accordingly: have consistant formatting, }

\subsection{Data Preprocessing}

\textbf{Data aggregation.} 
Usually, base stations periodically collect the associated user-level KPIs, which can be further processed for downstream analysis, e.g., failure prediction. Thus, we aggregate user-level time series KPIs to base station-level KPIs. Specifically, the total data collection time is divided into one-second time intervals. At each time interval, we aggregate the KPIs from UEs at the associated base station. This procedure gives the time series KPI data for discrete time intervals for all base stations in the network.
%\textcolor{red}{this sentence is not clear to me --- We perform an average aggregation over the time intervals to get time series KPI data for each time interval for the UEs. }  
%\textcolor{blue}{The data gets captured for different UEs for the time intervals to get all UE data for each of those one-second intervals. We perform an average aggregation over the time intervals to get time series KPI data for each time interval for the UEs.}
%At each time interval, we take the average aggregation of KPI data across all UEs a base station serves.  
%We average across sectors, which gives the time series base station KPI data.

\begin{figure} [ht]
    \centering
    \includegraphics[width=.8\columnwidth]{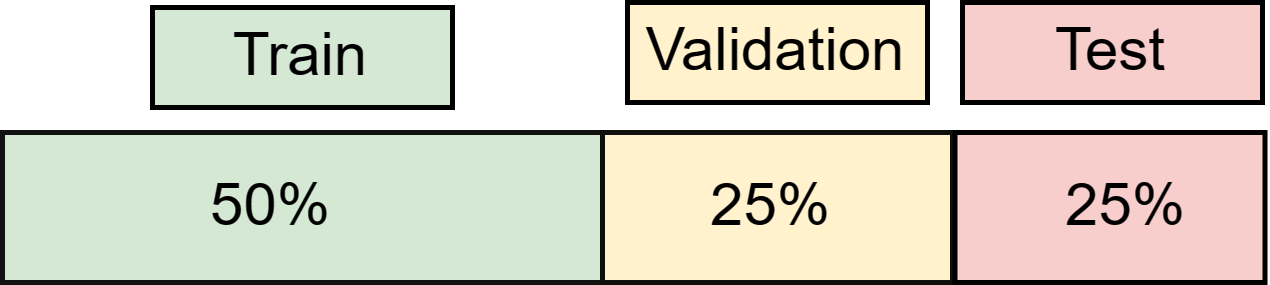}
    \caption{The time series data split.}
    \label{fig:rolling}
\end{figure}

\textbf{Handling data imbalance.} The percentage of failures, however, is less compared to the normally operating ones, which creates imbalanced data to be processed. We also have different percentage of failures for EPR and Interference. We use the weighted cross-entropy loss function to handle the data imbalance by combining prior class probabilities into a cost-sensitive cross-entropy error function. Other approaches such as SMOTE oversampling and under-sampling, has drawbacks like noisy sample generation and underutilized dataset. Unlike traditional cross-entropy, this approach results in well-balanced classifiers given different imbalanced scenarios while minimizing overall error. 
We calculate the class specific ratios by taking the number of samples for one of the failures and dividing it by the total number of samples. The loss function sets these prior class specific ratios as weights into the regular cross entropy (Eq.~\ref{eq:ce}). This ensures that all classes have an equal influence on the loss minimization.
%\textcolor{red}{The loss function sets the prior minority to majority  class ratio $\lambda = 0.03$ into the regular cross entropy (Eq.~\ref{eq:ce}).}\textcolor{red}{Hasan recheck this part --- This $\lambda$ ensures that both classes have an equal influence because when $y=0$ for a non-failure instance, the remaining term $(1-y^i)\log(1-\hat{y}^i)$ only contributes $\lambda= 0.3 $ percent to the loss. Similarly, when $y=1$ for a failure instance, the remaining term $-y^i\log(\hat{y}^i)$ contributes $(1-\lambda)=99.7$ percent to the loss.}
%\textcolor{blue}{ }

%\textcolor{red}{\begin{equation} \label{eq:ce}
%J(\theta) = \frac{1}{n} \sum_{i=1}^n -y^i log(\hat{y}^i) (1-\lambda) - (1-y^i) log(1-\hat{y}^i)\lambda
%\end{equation}}

\begin{equation} \label{eq:ce}
J(\theta) = \frac{1}{n} \sum_{i=1}^n -w_y log\frac{exp(\hat{y}^i)}{\sum_{c=1}^C exp(\hat{y}^i_c)}
\end{equation}
Here $J(\theta)$ and n are the total loss and number of samples respectively. $w_y$l is the class specific weight for ground truth class $y$, $C$ is the total number of classes. Lastly, the numerator of the logarithm refers to exponential of the predicted probability score by the model and the denominator is the sum of exponential of probability scores for all the other classes as predicted by the model.
\textbf{Time series split.} 
In the considered 5G RAN environment, the properties (e.g., KPIs) of base stations evolve over time, i.e., the assumption of data with independent and identical distribution is invalid. In such cases, the classical regression and classification model assessment approach of using cross-validation is not the right fit. Thus, we perform time series split into the data by sorting it across time and splitting them into the first 50\% train, the following 25\% validation, and the last 25\% test set to create a split with the largest training data. The time series data split is presented in Fig. \ref{fig:rolling}. 

%Cross-validation (CV), a widely adopted method for assessing algorithm generalizability in classification and regression problems. Our dataset contains time series KPI features and the underlying process evolves over time. Due to this reason, the process can undermine the fundamental assumptions of cross-validation that the data is independent and identically distributed. Therefore, we use \textit{rolling origin} \cite{kreindler2016effects} method to handle the temporal dependency and patterns while comparing our framework with previous works. 

%\textcolor{blue}{ The process involves sequentially moving values from the validation set to the training set while changing the forecast origin and including more train data accordingly. This is also known as \textit{n-step-ahead} evaluation, where $n$ is the forecast horizon. We create the 5 folds by first sorting the data across time and splitting them into the first 70\% train, the next 20\% validation and the last 10\% test set to create the fold with the largest training data. Subsequent folds are created by offsetting the splits by 10\% such as the second fold would contain the first 60\% as the train, the next 70\%-80\% as the validation, and the next 80\%-90\% as the test data. (Fig. \ref{fig:rolling}). Similarly, we create the rest of the folds for both deployments.}

\subsection{Model Architecture}

\begin{figure*} 
    \centering
    \includegraphics[width= \textwidth]{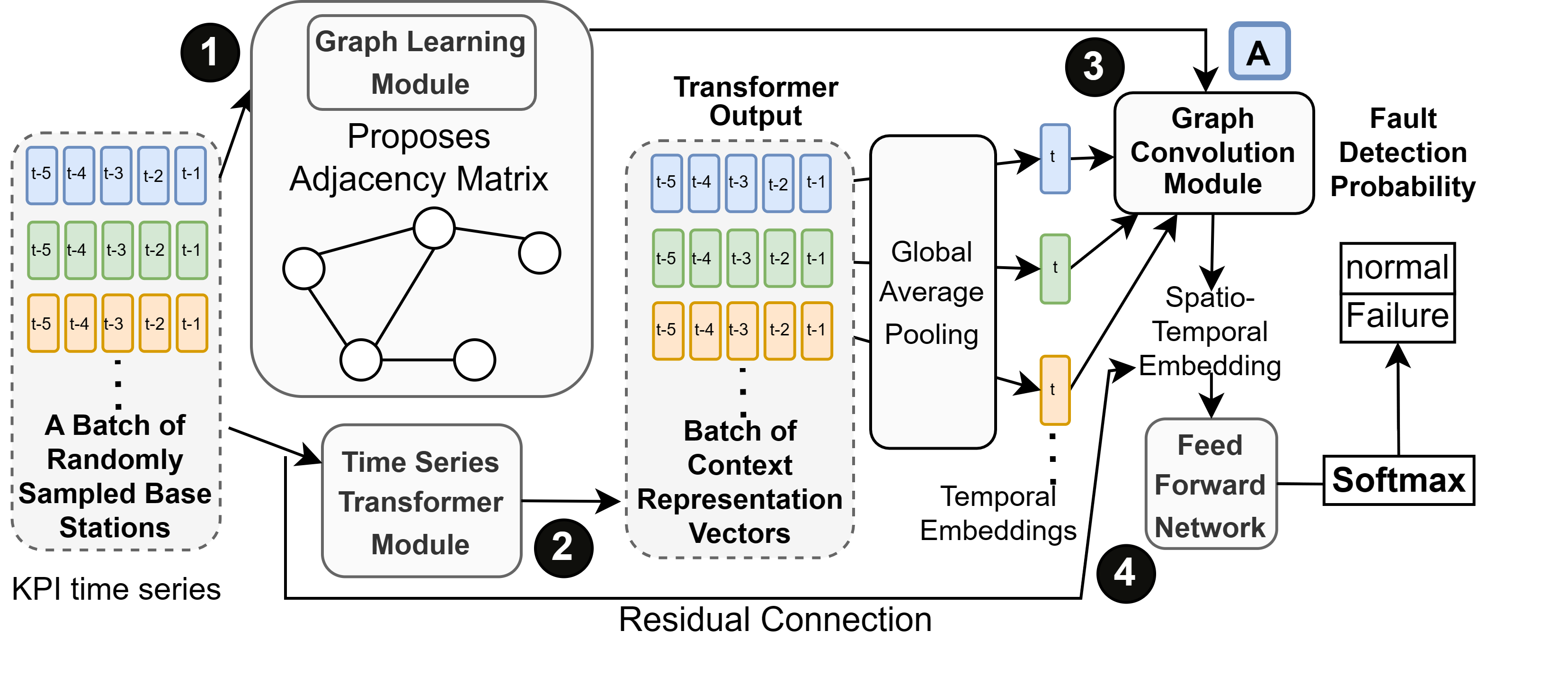}
    \caption{The architecture of Simba.}
    \label{fig:model}
\end{figure*}

Fig.~\ref{fig:model} presents the overall architecture, which maps base station time-series data to a probability distribution over anomalies. Specifically, the prediction system takes the pre-processed time series KPIs data as inputs and generates the output as a vector of the probability distribution of different anomalies for the next time step. The system deploys \circled{1} a GL module to learn a graph structure from unstructured time series data, \circled{2} a time series Transformer module with global pooling for temporal feature embedding, and \circled{3} a GC module to generate spatial feature embedding of the graph structure generated by GL module. The spatial embedding from the GC module and the temporal embedding from the Transformer module are then aggregated to generate spatio-temporal embedding, which \circled{4} a feed-forward network processes to generate the final output. The output vectors are real numbers that a Softmax function normalizes as the probability distribution of different anomalies. We divide Simba architecture into three modules: GL and GC modules to capture spatial features, a time series transformer for temporal features, and a feed-forward network for classification and generating the output vector. Each module is elaborated below.  

\textbf{GL and GC modules.}
In unstructured time series KPI data from a collection of base stations, the change in a base station's KPI may cause changes in neighboring base stations' KPIs due to their propagation relationship in radio space. The graph learning layer \cite {wu2020connecting} is designed to extract such relationships from the unstructured time series data (KPIs). Specifically, GL can learn stable relationships among base stations (nodes) throughout the training. As the model gets trained, the GL module better proposes the optimal graph structure for the training dataset. The GL module can be described using the following equations:
%\textcolor{red}{ Once the model is trained in an on-line learning version, the graph adjacency matrix is also adaptable to change as new training data updates the model parameters.} \textcolor{blue}{} 

\begin{equation} \label{eq:gsl1}
M_1= tanh(\alpha E_1 \theta_1)
\end{equation}
\begin{equation} \label{eq:gsl2}
M_2= tanh(\alpha E_2 \theta_2)
\end{equation}
\begin{equation} \label{eq:gslA}
A= ReLu(tanh(\alpha ( M_1M_2^T-M_2M_1^T ))
\end{equation}
% was blue before here
Here, $E_1, E_2$ denotes node KPI features; $\theta_1$ and $\theta_2$ are the neural network parameters for linear layers. $\alpha$ is a control variable for the saturation rate of the activation function. Equation \ref{eq:gslA} provides the adjacency matrix with edge weights that denote the degree to which one node influences another. For each node, the top $k$ closest neighbors are then selected, and the remaining neighbors' distances are set to zero, which provides the final graph structure achieved from the module. 

\begin{figure} [ht]
\centering
\subfloat[GC layer]{\includegraphics[width=.47\linewidth]{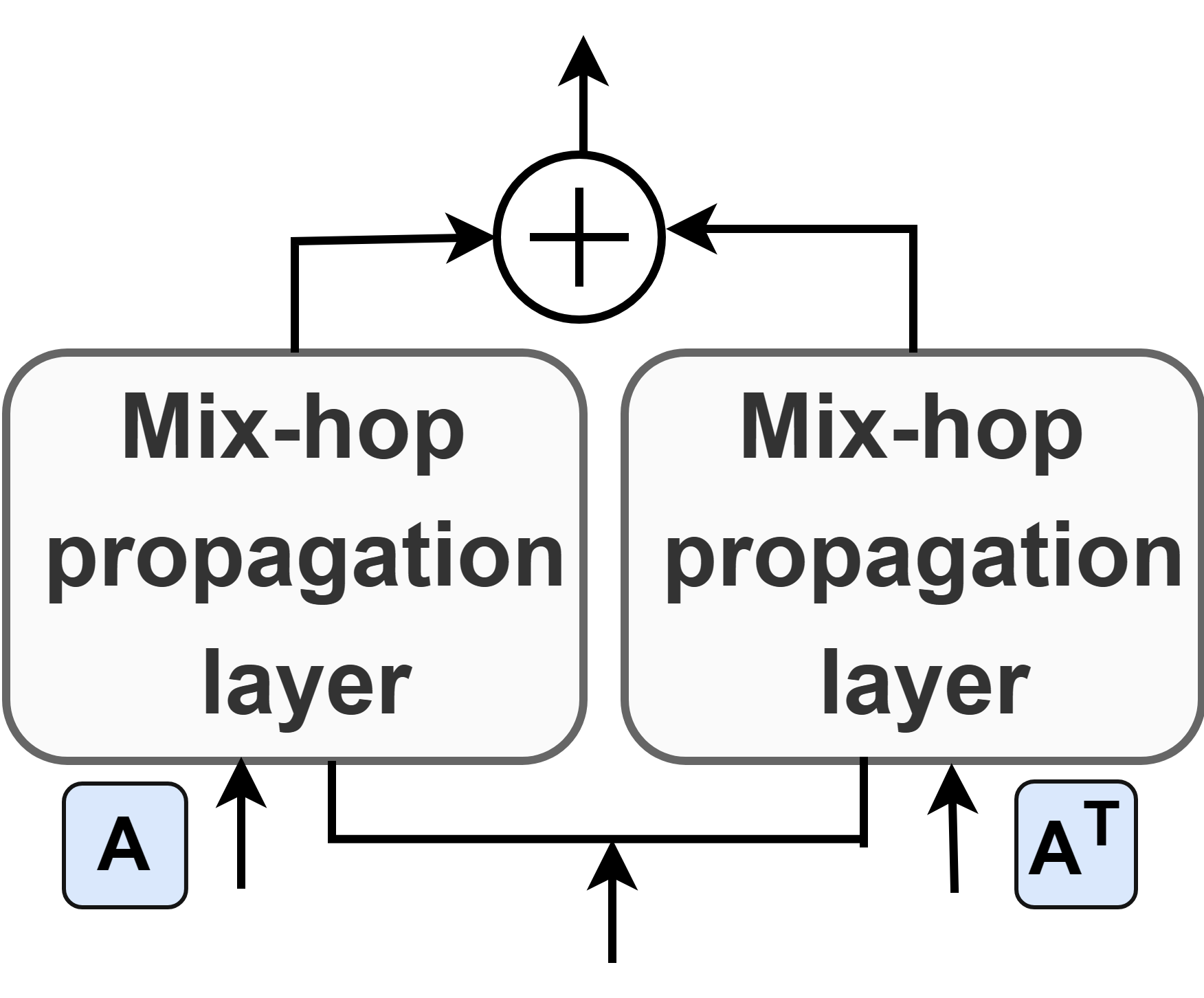}}
  \quad
\subfloat[Mix-hop propagation layer]{
  \includegraphics[width=.47\linewidth]{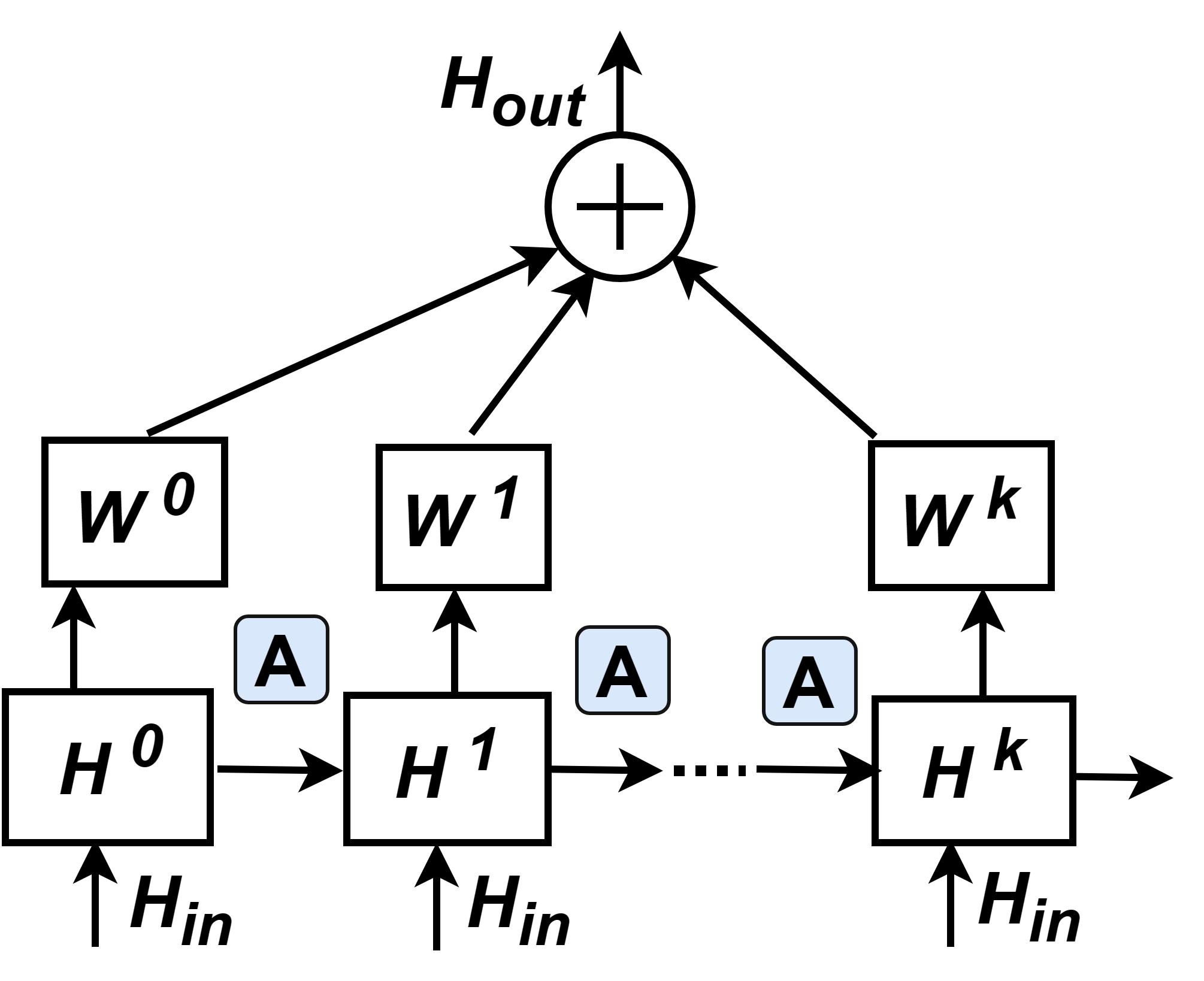}}
\caption{GC module and mix-hop propagation layer.}
\label{fig:gc}
\end{figure}

%\textcolor{red}{Hasan pls read the following para again to make sure we are on the same page.} \\
%\textcolor{blue}{
The graph convolution module aims to combine base stations' temporal embedding vectors with their neighborhood embedding vectors. We adopt the GC module developed in MTGNN \cite {wu2020connecting} (see Fig.~\ref{fig:gc}) that consists of two mix-hop propagation layers that process inflow and outflow information passed through each base station separately. The net inflow information is calculated by adding the outputs of the two mix-hop propagation layers. The mix-hop propagation layer consists of two steps: 1) the information propagation step responsible for the recursive information propagation of node and graph structure and 2) the information selection step that filters out useful information at each hop. 
%}
%\textcolor{red}{The information propagation step propagates node information along with the given graph structure recursively. A information selection step is introduced to filter out important information produced at each hop to discard useless information.}

\textbf{Transformer module.}
The transformer module extracts high-level temporal features of the KPI data. Initially, it receives time-series KPI data for the previous five time steps. The positional encoding scheme of the transformer encodes each position or index into a vector. Hence, the output of the positional encoding layer is a matrix, where each row of the matrix represents an encoded object of the sequence summed with its positional information. The transformer module has a multi head attention with 4 heads, each of size 32, and the two 1D convolution filters are of size 32 and 17, respectively (Fig.~\ref{fig:trans}). The transformer module converts the time series data into context representation vectors. We perform global average pooling across the time dimension vector to capture temporal dependencies for each base station. Thus, the transformer module generates one feature vector to capture the temporal features.

\textbf{Feed forward networks.}
The output vector from the convolutional module is concatenated with the output vector from the generalized transformer branch to get a representation vector that captures both temporal and static dependencies. We feed the concatenated vector to a feed-forward network with two layers: 16 and 2 neurons. A Softmax layer gets the final probability vector for fault prediction. 

\textbf{Model training} 
 There are hyper parameters that need to be learned during model training. These hyper-parameters (Table ~\ref{simba-hp}) are chosen based on problem-specific needs and parameter tuning experiments. Using a small batch size (e.g., 32 or 64) can generate unstable loss curves when extremely low minority-to-majority class ratios are present in datasets. Thus, we use a batch size of $2000$ for the deployment to ensure that each batch has at least $2$ failure events on average, which provided stability in model training. We predict the failure probability of a specific failure for the next instance using the KPIs from the five previous instances, offering the best prediction performance. The subgraph size is the same as the number of base stations in the deployment, which is 7. This ensures that we learn the influence of every base station on all the remaining ones.
 
\begin{table}[t]
    \centering
    \begin{tabular}{|p{4.2cm}|p{2.2cm}|} \hline
    
\textbf{Parameters
}& \textbf{values}    \\ 
\hline
Batch size & 2000 \\
\hline
Subgraph size for mix-hop & 7 \\
\hline
convolutional kernel size & 3 \\
\hline
number of convolution filters & 32 \\
\hline
GL module alpha & 0.5 \\
\hline
Mix hop alpha & 0.5 \\
\hline
Learning rate &  0.0003  \\
\hline
Convolution Kernel Size  &  3  \\
\hline
Number of convolution filters & 32  \\
\hline
Number of previous seconds as input & 5 \\ 
\hline
\end{tabular}
\caption{The list of hyper parameters in Simba model. }
\label{simba-hp}
\end{table}
 
 % hyperparameters: subgraph size - 7, convolutional kernel size - 3, number of convolution filters - 32, train size - 0.5, validation size - 0.25, test size - 0.25, learning rate - 0.0003}\textcolor{red}{it needs little more elaboration -- Hasan can you elaborate it please? also, need to list the hyperparameters. }

%\textcolor{blue}{ Using a small batch size (e.g., 32 or 64) could generate unstable loss curves when extremely low minority-to-majority class ratios are found in data sets. Thus, we use a batch size of $1024$ for rural deployment and $6000$ for urban deployment to ensure that each batch has at least $2$ failure events on average, which provided stability in model training. We predict the of failure probality of a specific failure for next day using the previous $5$ days' KPIs which offers the best prediction performance as stated in paper \cite{islam2022deep}).}

%%%%%%%%% Let's put existing architecture in a seprate section

\section{State-of-the-art Architectures}

This section presents two state-of-the-art architectures: GNN\_RCA \cite{yen2022graph} and MTGNN \cite{wu2020connecting} that build on GNN for spatial and temporal context capture. We will first describe their architectures for the comparison with Simba and then state their gaps that Simba fills.   

\begin{figure*} [ht]
    \centering
    \includegraphics[width=0.9\textwidth]{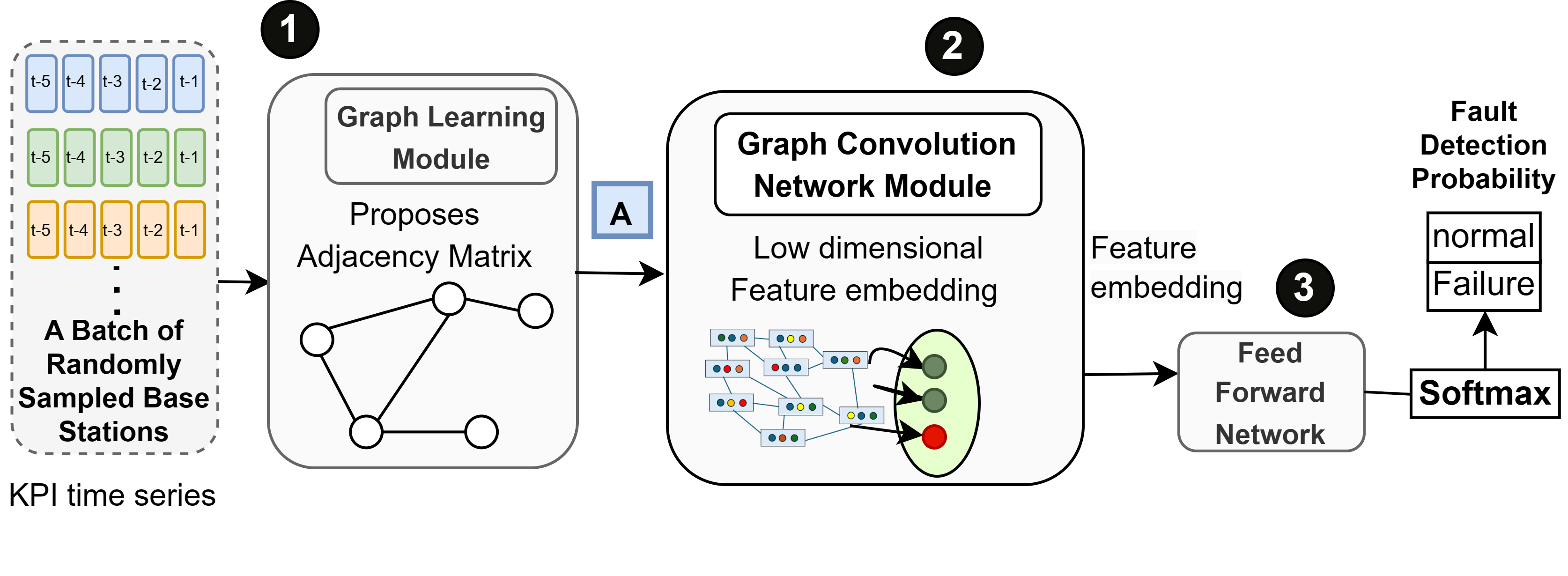}
    %\includesvg[width=0.4\textwidth]{pictures/automation-workflow.drawio.svg}
    \caption{GNN\_RCA Architecture.}
    \label{fig:gnn_rca}
\end{figure*}

\subsection{GNN-based anomaly detector (GNN\_RCA)}

The architecture of GNN\_RCA is depicted in Fig.~\ref{fig:gnn_rca}. We follow the same data pre-processing steps as used in Simba. As in Simba, the input represents the feature vectors of base stations' time-series KPI data,  which we feed into \circled{1} Graph Learning (GL) module. It takes this unstructured time series KPI data and generates graph structure using Equation~\ref{eq:gsl1} -~\ref{eq:gslA}. An adjacency matrix $A$ is initially generated using top $k$ (e.g., 5 in our experiments) weights as edges, where the weights infer the likelihood of connections among nodes (base stations). This weighted adjacency matrix represents the learned graph structure required as input to the next module \circled{2} GCN. It initially uses a linear transformation to transform the input features of each node into a low-dimensional representation. Then, message passing and graph convolution operation is performed on that low-dimensional feature vector between a node and its neighbours to generate the learned feature embedding using Equation~\ref{eq:gcn}. The aggregation and message passing continues until all nodes have their optimal embeddings.   

The learned optimal embeddings are then fed into a feed-forward network \circled{3} to generate the final output vector that a Softmax function normalizes as the probability distribution of different anomalies. Note that GNN\_RCA learns spatial feature embeddings at each time instance without incorporating the temporal context of the features, which may impact the model's effectiveness in the considered time series KPIs. We train the developed model following the same procedure as in Simba, and list the learned hyperparameters in Table~\ref{simba-hp}. 

% \begin{table}[t]
%     \centering
%     \begin{tabular}{|p{4.2cm}|p{2.2cm}|} \hline
    
% \textbf{Parameters
% }& \textbf{values}    \\ 

% \hline
% Top neighbors $k$  &  5   \\
% \hline
% Batch size &  2000 \\
% \hline

% Saturation rate, $\alpha$ &  0.5-5  \\
% \hline
% Learning rate, $\sigma$ &  0.0003  \\
% \hline
% Convolution Kernel Size  &  3  \\
% \hline
% Number of convolution filters & 32  \\
% \hline
% Epochs &  140 \\
% \hline
% \end{tabular}
% \caption{The list of hyper parameters in GNN\_RCA model. }
% \label{gnn-rca-hp}
% \end{table}

%batch size: 2000, 2000, 2000 Learning rate: 0.0003, 0.0003, 0.0003 Past time point data: 1s, 5s, 5s Epoch: 140, 140, 140, 140
%is applied for node features embedding of the graph. The GCN contains multiple neural layers to get the optimal embeddings of nodes' KPIs using Equation~\ref{eq:gcn}. which are then finally mapped onto one feature vectors ( referred to as low-dimensional embedding in the paper). This is the spatial embedding for the nodes which \circled{3} a feed forward network processes to generate the final output vectors that a Softmax function normalizes as the probability distribution of different anomalies. As this model does not handle the temporal features separately, it might not be able to perform well for the KPI data as per our expectation.
%\textcolor{red}{add the list of hyper parameters in a table. did we learn the value of k? if yes, then it will be in the hyperparameters.}
%\textcolor{red}{Hasan: check if this description needs including equations, etc., to make things much clearar for novice readers. } \\

\subsection{Multi Variate Time-series Graph Neural Network (MTGNN)}

\begin{figure*} [ht]
    \centering
    \includegraphics[width=0.9\textwidth]{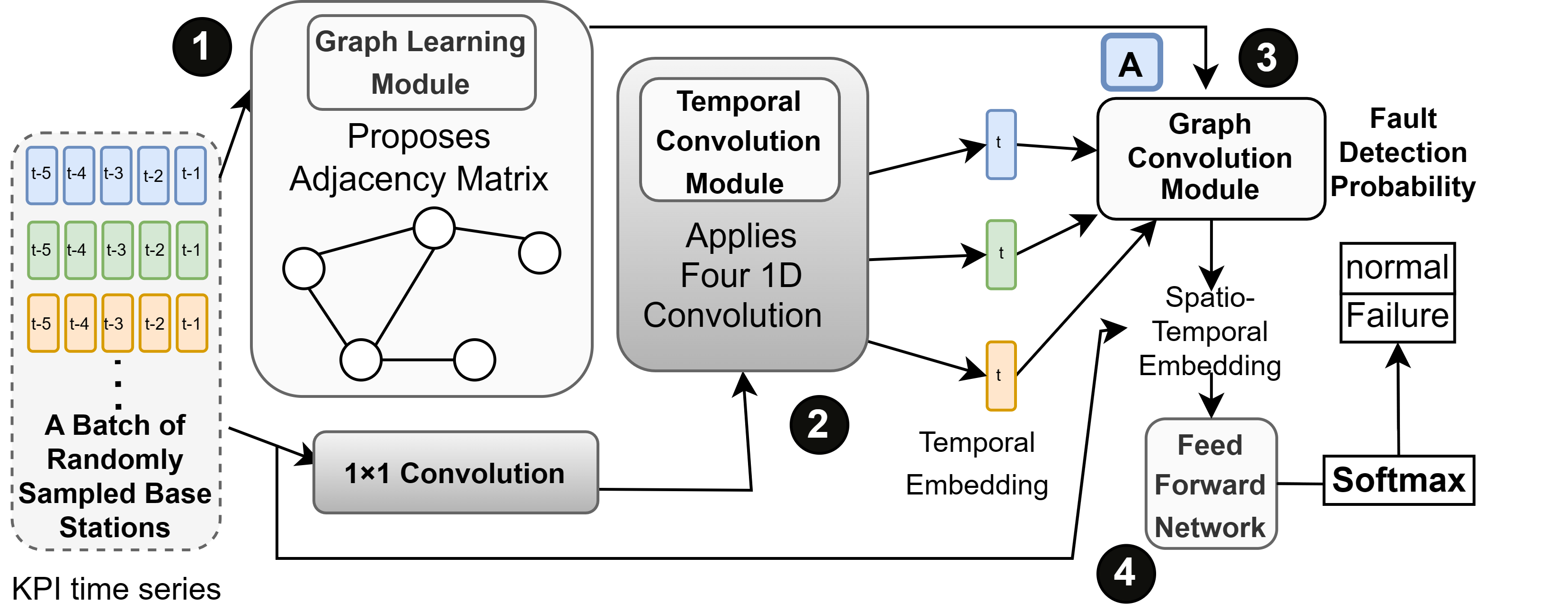}
    \caption{ MTGNN architecture. }
    \label{fig:mtgnn}
\end{figure*}

The architecture of the MTGNN model is shown in Fig.~\ref{fig:mtgnn}. This model takes the same preprocessed data used in the previous two models. Specifically, we feed the base stations' time-series KPI data as input into \circled{1} Graph Learning (GL) module as in Simba and GNN\_RCA. It generates the graph structure following the same equations. To develop the weighted adjacency matrix $A$, we use the same value of $k$ here. In parallel, our KPI input data are fed into a $1 \times 1$ standard convolution, which projects the inputs into a latent space to fit into \circled{2} temporal convolution (TC) module. In the TC module, different standard $1D$ convolution filters like $1 \times 2$, $1 \times 3$, $1 \times 6$, and $1 \times 7$ are used, which finally generate one temporal feature for each node (temporal embedding).

%We also implement the MTGNN \cite{wu2020connecting} approach to compare it with our Simba and other models. We also follow the same data preprocessing steps as mentioned earlier for our model. Figure \ref{fig:mtgnn} represents the complete architecture of the MTGNN model.As in Simba and GNN\_RCA, we feed the base stations' time-series KPI data as input into \circled{1} Graph Learning (GL) module. It generates the graph structure following same Equation~\ref{eq:gsl1} -~\ref{eq:gslA}. For generating the weighted adjacency matrix $A$, we use the same value of $k$ here. In parallel, our KPI input data are fed into a $1 \times 1$ standard convolution which projects the inputs into a latent space to fit into \circled{2} temporal convolution (TC) module. In TC module, different standard $1D$ convolution filters like $1 \times 2$, $1 \times 3$, $1 \times 6$ and $1 \times 7$ are used which finally generates one temporal feature for each node (temporal embedding). 

The weighted adjacency matrix from GL module and temporal embedding from TC module are fed into \circled{3} graph convolution (GC) module which follows the similar procedure as our model's GC module. First, GC module collects nodes’ neighborhood information from $A$ and generates spatial feature embedding of the nodes. Then, it combines the spatial features with the temporal feature embedding and generates spatio-temporal features for the base stations. Residual connections are added from the inputs of temporal convolution module to the outputs of GC module to avoid the problem of gradient vanishing. The rest of the procedure is similar to other models like using \circled{4} a feed forward network and softmax to generate probability distribution of different faults from the spatio-temporal embeddings. This model handles both spatio-temporal features of the KPI data, however, it only uses 1D convolution filters in TC module which may not be enough for the long term time series data. We train the developed model following the same procedure as in Simba and used the same hyperparameters listed in Table~\ref{simba-hp} for GNN\_RCA.

\section{Results and Discussion}

This section first presents the evaluation setup and the chosen performance metrics. Then, we present the performance comparison of Simba with the state-of-the-art MTGNN-based model in the chosen deployment scenario by answering the following research questions related to our objective and contributions. 
%We conclude this section with the final takeaway and extension possibilities of Simba. 

\begin{itemize}       
\item \textbf{RQ1 (model comparison)}: Is Simba an effective model for anomaly detection and root cause analysis compared to state-of-the-art ones? 

\item \textbf{RQ2 (failure comparison)}: How Simba behaves on the chosen faults? 

\item \textbf{RQ3 (user density comparison)}: Is Simba effective on varying user densities that define operational hours like peak and normal? 
\end{itemize}

\subsection{Performance Metrics and Evaluation Setup}

We evaluate the performance of different models using three metrics: precision, recall, and F1-score. For each approach, we first calculate true positives (TP), true negatives (TN), false positives (FP), and false negatives (FN) cases for all failures and the normal case. \textit{True positives} are outcomes where the model correctly predicts the positive classs (failure events for this data set). Similarly, \textit{True negatives} are outcomes where the model correctly predicts the negative class (non-failures events). \textit{False positives} are the negative events (non-failures) that are predicted as positive (failures), while \textit{False negatives} are positive (failures) events that are predicted as negative (non-failures).
\textit{Precision} a measure of how many of the positive predictions made are correct (true positives) while
\textit{Recall} quantifies the measure of positive class predictions made out of all positive examples in the dataset.
\textit{F1 Score} provides a single score that balances both the concerns of precision and recall in one number. It is generally described as the harmonic mean of the two values.
We calculate the metrics for both failure and non-failure classes as follows:
\begin{equation}
Precision = \frac{TP}{TP+FP}
\end{equation}
\begin{equation}
Recall = \frac{TP}{TP+FN}
\end{equation}
\begin{equation}
F1_{score} = \frac{2×Precision×Recall}{Precision+Recall}
\end{equation}

We note the precision, recall, and F1-score of different failure and non-failure .
We run the experiments on Intel(R) Xeon(R) Silver 4210R CPU with 2.40GHz and 32 GB memory, and Nvidia Quadro RTX 8000 GPU with 50GB VRAM. We use Ubuntu 20.04.6 LTS OS and CUDA 11.7 GPU versions, respectively.

\subsection{Performance Comparison}

%\textcolor{blue}{two deployments: Chicago vs. GPPP. Chicago: we have normal vs. peak.  3 models. 3 faults.}
%For our performance comparison we choose two deployments: Chicago and GPPP deployments where Chicago has 7 and later one contains 19 base stations. We train our \textit{Simba} model along with other two models on different training folds and report performance on the 5-fold test data.
% The F1-scores with their corresponding precision and recall are shown in Table \ref{table1}. The result represents the performance of different models for the 3GPP deployments with 7 base stations, in a peak (90 users per base station) and normal demand (30 users per base station) scenario. All these results are for failure prediction for next time step. 

\textbf{Model comparison (R1):} In the presence of the chosen faults, we present the results of the above three models on both normal and peak operations in Table~\ref{tab:model-comp}. We observe that \textit{Simba} significantly and consistently outperforms all existing approaches, achieving the best F1 scores for both normal and peak hours of operation. Specifically, in the case of normal operation, \textit{Simba} performs the best in terms of precision and recall with scores of 99\% and 95\%, respectively. During peak hours, Simba still has the highest precision and F1 score, with a slightly worse recall than MTGNN. Out of the three models, GNN\_RCA performed the worst in terms of F1 score in both scenarios. Also, it has the worst precision in normal operation (76\%) and the worst recall during peak hours (75\%).
% We can observe that \textit{Simba} significantly and consistently outperforms all existing approaches, achieving the best F1 scores consistently for all types of failures and in both normal operation and peak hours. Our Simba model scores over 96\% F1 scores both for correctly identifying normal activity and Excessive Power Reduction (EPR) failure on both scenarios. On the other hand, on the peak scenario, MTGNN has 93\% F1 score for both cases while GNN\_RCA performs worst achieving 88\% for EPR as well as 80\% for the other one respectively. This behavior is similar on the normal operation scenario, with MTGNN achieving 96.3\% and 98\% for EPR and no-failure prediction, and GNN\_RCA achieving 96\% for both.
% Additionally \textit{Simba} has a F1 score of 84\% and 95\% for interference in the peak a normal operation scenarios respectively. MTGNN and GNN\_RCA scores 82\% and 68\% respectively in the peak scenario and 85\% and 69\% in the normal operantion scenario. 
%For identifying Too Late Hand Over (TLHO), all the models show performance degradation among which our model secures 20\% F1 scores compared to other models.

The inferior performance of GNN\_RCA can be attributed to the fact that it can not handle temporal features of the KPI data using only the GCN module, which leads to the performance degradation of the model than others. Moreover, it uses a simple GCN module compared to the other two models, failing to filter out important feature information produced at each node. GNN\_RCA simply aggregates all neighbourhood information that sometimes may add unnecessary information or noise to the feature vectors.

On the other hand, although MTGNN uses a GC module similar to the GNN\_RCA for spatial feature extraction, its Temporal Convolution (TC) module for temporal feature handling considers equal weights for all previous time step data generating performance loss compared to our model. This TC module does not have any internal mechanism to give higher priority to essential events, e.g., feature values from the most recent events. Because of this limitation, this model fails to capture complex dependencies in a long-span sequence. In the case of the \textit{Simba} model, we use the GC module for spatial and the transformer for temporal feature extractions. Transformer uses a self-attention mechanism to focus on the most relevant elements of the input sequence and the complex relationships among feature values that are far apart from each other in the sequence. These advantages from transformer time series lead to the superior performance of \textit{Simba}.

\begin{table}[t]
    \centering
    \resizebox{\columnwidth}{!}{%
    \begin{tabular}{|c|c|c|c||c|c|c|} \hline
    & \multicolumn{3}{c||}{Normal Operation} & \multicolumn{3}{c|}{Peak Operation} \\ 
    \hline
    \textbf{Model} & \textbf{Precision} & \textbf{Recall} & \textbf{F1} & \textbf{Precision} & \textbf{Recall} &  \textbf{F1}    \\ 
    \hline
    \textbf{GNN\_RCA} & 0.76 & 0.95 &  0.82 & 0.82  & 0.75 & 0.78 \\
    \hline
    \textbf{MTGNN} & 0.95 & 0.87 & 0.91 &  0.79 & 0.99 & 0.88 \\
    \hline
    \textbf{Simba} & 0.99 & 0.95 &  0.97 &  0.89 & 0.92 & 0.90\\
    \hline
    \end{tabular}
    }
    \caption{Performance comparison for different models. }
    \label{tab:model-comp}
\end{table}

\textbf{Failure comparison (R2):} We also observe a substantially high performance of Simba compared to the two studied failures, depicted in Table \ref{tab:fault-comp}. In particular, Simba performs the best in the case of the Excessive Power Reduction (EPR) fault with near-optimal precision, recall and F1 score in normal operation hours. Similarly, it offers high performance in peak hours, confirming its effectiveness in various operating conditions. The reduced transmission power in the affected station results in a drop in KPIs like RSRP and RSRQ. Consequently, the coverage of the affected cells is drastically reduced, leading to a decrease in the number of users connected to that base station. These consequences of EPR create a prominent change in the corresponding KPIs, leading to a high performance of Simba.  

Interference presents a slightly more subtle effect, which leads to a decreased performance for Simba compared to EPR. Specifically, Simba achieves an F1 score of 0.95 and 0.84 for normal and peak hours, respectively. In the peak operation hours, the precision slightly drops, meaning that the model misclassified some of the normal samples as interference ones. This outcome is because the interference effect is more nuanced compared to EPR, where we see degradation in SINR and signal quality. However, such a change is not as pronounced and abrupt as is the case of EPR. Consequently, we observe a slightly higher misclassification. However, Simba still achieves high scores in terms of recall, rarely overlooking interference samples.

\begin{table}[t]
    \centering
    \resizebox{\columnwidth}{!}{%
    \begin{tabular}{|c|c|c|c||c|c|c|} \hline
    & \multicolumn{3}{c||}{Normal Operation} & \multicolumn{3}{c|}{Peak Operation} \\ 
    \hline
    \textbf{Model} & \textbf{Precision} & \textbf{Recall} & \textbf{F1} & \textbf{Precision} & \textbf{Recall} &  \textbf{F1}    \\ 
    \hline
    \textbf{EPR} & 0.99 & 0.99 &  0.99 & 0.99  & 0.92 & 0.96 \\
    \hline
    \textbf{Interf} & 0.99 & 0.91 & 0.95 &  0.78 & 0.91 & 0.84 \\
    \hline
    \textbf{No Failure} & 0.98 & 0.99 &  0.98 &  0.99 & 0.93 & 0.96\\
    \hline
    \end{tabular}
    }
    \caption{Simba performance comparison for different failures. }
    \label{tab:fault-comp}
\end{table}

%Finally, Too-Late Handover is the one failure that all the models had the most difficulty with, as Simba had the highest performance achieving a F1 Score of 0.20. This can be explained due to intrinsic nature of the fault itself. Differently from both EPR and Interference, TLHO does not immediately affect the KPIs when it occurs. As the only change made is to change the threshold for performing a handover, the effects of it can only start to be seen when users try to move to a different cell. Even when that occur, the effect is subtle, as the failure does not abruptly reduce the strength of signals for users, it simply allows the signal to gradually worsen. Moreover, this fault is largely subjected to the randomness of the simulation, as users move through space utilizing a Random Waypoint mobility model, where a base-station can be suffering from TLHO but because most of the users connected to it do not move in the direction of another cell, no large effect can be seen in the aggregated KPIs for the station.  

\textbf{User density comparison (R3)}: As seen in Table \ref{tab:fault-comp}, Simba is effective in both normal and peak-hour operations. Specifically, it performs slightly better on the normal operations compared to the peak-hour operation. For instance, its recall in peak-hour operation is slightly worse for both EPR and non-failure events. However, in the case of Interference, Simba's recall is similar in both scenarios. 
%\textcolor{red}{need a little more clarification  --- However, the proposed architecture performs slightly better for both EPR and Interference in the normal operation case. We can see that for both EPR and the No Failure case, the precision achieved by Simba is high in both scenarios, but the recall is slightly worse during peak hours.}

The behavior experienced with the EPR fault can be explained by the aggregating method utilized. As we average out the users' KPIs to calculate a single base-station measurement, a larger number of users can end up diluting the effect of failures after aggregation due to some users being in a position where they are less affected by the fault. This can have the effect of masking EPR cases, leading to a slightly worse recall.

For the Interference behavior, the reduction in precision indicates that, during peak operation, Simba classifies more non-failure samples as suffering from the failure. This drop in precision can be explained first by the already mentioned subtlety of this fault as well as the increase in the number of users, leading to more varied performance from users operating without any failures. With the increased density, we tend to see more non-faulty users in worse conditions that can be misclassified as suffering from Interference, which leads to a slight drop in precision.

\section{Conclusion}
In this work, we leveraged a calibrated simulator to generate normal and faulty data for two different scenarios, peak and off-peak demand. Utilizing this data for training, we proposed a novel architecture, \textit{Simba}, capable of dealing with the spatial-temporal complexity of telecommunication networks, and perform anomaly detection and Root-Cause Analysis. Our solution utilized state of the art GNN and Transformer models, integrating them for a novel architecture. We compared our model results to two other state-of-the-art approaches, MTGNN and GNN\_RCA, and found that our model outperform both in detecting and labeling the failures studied.

While Simba is thoroughly evaluated for specific scenarios, the 3GPP default scenario focus on an Urban eMBB environment solely. Other 3GPP defined scenarios encompass different environments such as Rural and Indoor, and varied classes of use cases such as Ultra-Reliable Low Latency Communications (URLLC) and Massive Machine-Type Communications (mMTC). Further research could possibly explore these different configurations to expand the evaluation conducted with Simba.

%In future works, we intend to extend our work by testing on more configuration scenarios, such as Rural or Indoor, and integrate our tool to a real testbed, thus adapting our system to fully work with real-world data.

\ifCLASSOPTIONcaptionsoff
  \newpage
\fi
\bibliographystyle{IEEEtran}
\bibliography{reference}


%\newpage

% \end{comment}


%\vfill


\end{document}